\documentclass[12pt]{article}

\usepackage{times}
\usepackage{graphicx}
\usepackage{color}

\newenvironment{sciabstract}{%
\begin{quote} \bf}
{\end{quote}}

\newcommand{\red}[1]{\textcolor{black}{#1}}

\title{Coherent perfect absorption of nonlinear  matter waves}

\author{Andreas M\"ullers$^1$, Bodhaditya Santra$^1$, Christian Baals$^{1,2}$, 	Jian Jiang$^1$,\\
  Jens Benary$^1$,	 Ralf Labouvie$^{1,2}$,
		 Dmitry A. Zezyulin$^{3}$,  \\
		  Vladimir V. Konotop$^{4\ast}$, and Herwig Ott$^{1\ast}$\\
\\
\normalsize{$^1$ Department of physics and OPTIMAS research center,}\\ 
	\normalsize{Technische Universit\"at Kaiserslautern,  Erwin Schr\"odinger Stra{\ss}e, }\\
\normalsize{ 67663 Kaiserslautern, Germany}\\
\normalsize{$^2$Graduate School Materials Science in Mainz, Staudinger Weg 9, 55128 Mainz, Germany }\\
\normalsize{$^3$ITMO University, St.~Petersburg 197101, Russia}\\
\normalsize{$^4$Centro de F\'{i}sica Te\'orica e Computacional and Departamento de F\'{i}sica,}\\
\normalsize{ Faculdade de Ci\^encias, Universidade
	de Lisboa, Campo Grande, Ed. C8,}\\
\normalsize{   Lisboa 1749-016, Portugal}\\
\\
\normalsize{$^\ast$To whom correspondence should be addressed; E-mail: vvkonotop@fc.ul.pt (V.V.K),} \\ \normalsize{ott@physik.uni-kl.de (H.O.)}
}

\date{}

\begin{document} 


\baselineskip20pt


\maketitle


\begin{sciabstract}
 Coherent perfect absorption is the complete extinction \red{of incoming} radiation by a complex potential in \red{a physical system supporting wave propagation}. The concept was proven for \red{linear waves in} variety of systems including light interacting with absorbing scatterers, plasmonic metasurfaces and graphene films, as well as sound waves.  We extend the paradigm to \red{coherent perfect absorption of nonlinear waves} and experimentally demonstrate it for matter waves in an atomic Bose-Einstein condensate. Coherent absorption of nonlinear matter waves is achieved easier than its linear analogs because the strength of two-body interactions offers additional freedom for control. Implementation of the coherent perfect absorber of Bose-Einstein condensates paves the way towards broad exploitation of the phenomenon in nonlinear optics, exciton-polariton condensates, acoustics, and other areas of nonlinear physics. It also opens perspectives for designing atomic laser. 
\end{sciabstract}

\noindent{\bf One Sentence Summary:} Eliminating atoms of a Bose-Einstein condensate in an optical lattice from one cell creates a coherent perfect absorber for the quantum liquid.


\section*{Introduction}

 \red{Coherent perfect absorption (CPA) is the complete extinction of incoming radiation by a complex potential embedded in a physical system supporting wave propagation. The phenomenon is based on destructive interference of transmitted and reflected waves.}  The concept was introduced~\cite{StonePRL} and observed experimentally~\cite{StoneScience} for light interacting with absorbing scatterers. CPA was also reported on for plasmonic metasurfaces~\cite{meta}, graphene films~\cite{graphene}, and sound waves~\cite{acoustCPA}. Technologically, CPA is used to design switching devices~\cite{switching}, logic elements~\cite{OptLogic}, in interferometry~\cite{StoneScience} and in  many other applications~\cite{review}. \red{All these studies deal with perfect absorption of linear waves.} Here, we extend the paradigm to \red{a   CPA of {\em nonlinear} waves} and experimentally demonstrate it for matter waves with an atomic Bose-Einstein condensate (BEC). Conditions for CPA of matter waves can be satisfied easier than for its linear analogs because the strength of two-body interactions offers additional freedom for control. The observation of CPA of nonlinear matter waves paves the way towards a much broader exploitation of the phenomenon in nonlinear optics, exciton-polariton condensates, acoustics, and other areas of nonlinear physics.

\begin{figure} 
	\centering
	
	\includegraphics[width=0.8\columnwidth]{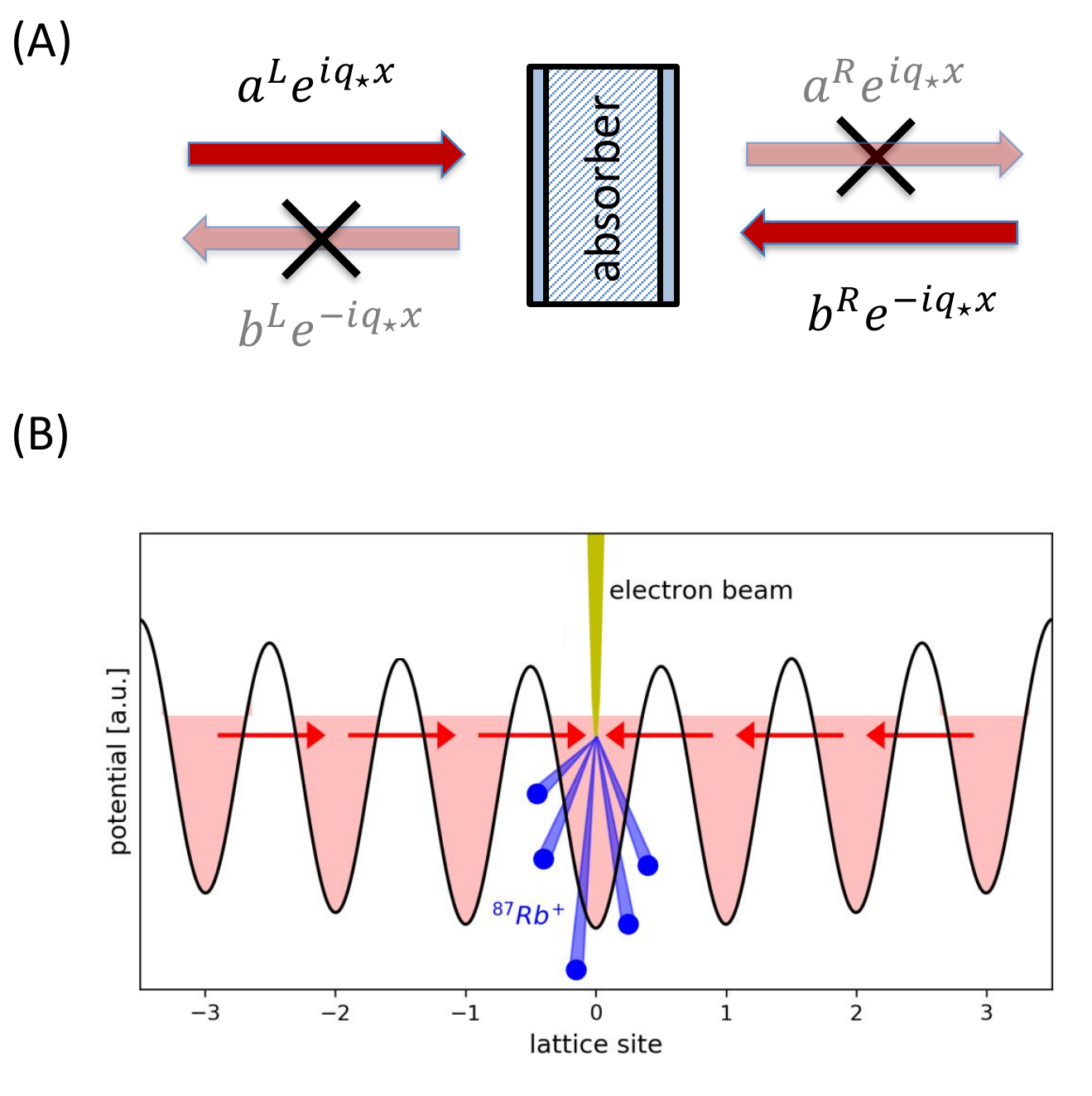}%
	\caption{
Working principle of coherent perfect absorption
(A) 
Incident radiation enters an absorber from two sides with a wavevector $q_\star$. Provided the complex amplitudes $a^L$ and $b^R$ are chosen properly, no radiation is transmitted or reflected. 	  	  	
(B) Experimental realization with a BEC  in an optical lattice. The matter waves enter a lossy lattice site from both sides. The losses are realized with an electron beam, which removes the atoms. Due to the interactions between the atoms, the two incoming waves are nonlinear, \red{while the absorption in the lossy site is linear.}  
}%
\label{fig:working_principle}
\end{figure}

CPA is a delicate phenomenon requiring precise tuning of the absorber and of the relative phases of the incoming waves. When the respective conditions are met for a particular wave vector, 
the radiation incident from both sides is completely absorbed 
(Fig.~\ref{fig:working_principle}A). CPA can be viewed as a time reversed process to lasing~\cite{StonePRL,StoneScience} where the absorber is replaced by a gain medium and only outgoing radiation exists for a given wave vector.  This time-reversed process is related to the mathematical notion of a spectral singularity~\cite{Neimark}, i.e. to a wave vector at which the system can emit radiation with none incoming~\cite{Mostafazadeh2009,Laser1}. Therefore, in the scattering formalism, the wavelength at which CPA occurs is called a time-reversed spectral singularity.

Recently, the concept of {\it nonlinear} CPA was introduced \red{in optics}~\cite{Mostafa2013,Gupta2013,Mostafa2014,Achilleos2016}. \red{ Such a device represents a nonlinear absorbing slab, sometimes with nontrivial composite internal structure~\cite{Gupta2014,Argyr}, which is embedded in a linear medium and thus perfectly absorbs incident linear waves.} 
Since, however, the propagating medium can itself be nonlinear, as it is the case for an optical Kerr nonlinearity or an interacting BEC, the natural (and still open) question arises about the physical meaning of CPA in a {\em nonlinear medium}. In other words, what are the scattering properties of nonlinear waves interacting with a linear absorbing potential?

The implementation of CPA in a nonlinear medium offers several challenges which raise doubts about whether such a phenomenon can exist and whether it is physically meaningful. Indeed, in such a setting the ``linear'' arguments do not work: there is no well defined transfer matrix connecting left- and right- incident waves (problems, where either the incident or transmitted radiation is given, have different solutions~\cite{DevSouil}); because of interactions among the modes there exists no interference in the linear sense, and results on scattering of monochromatic waves do not give any more an answer on the scattering of wave packets used in experiments.  Moreover, even if CPA can exist in a nonlinear medium, its realization is still questionable. There exists no general method of computing the system parameters, like the zeros of the transfer matrix elements in the linear case. Thus, tuning system parameters {\em in situ} might be the only possibility to realize CPA in nonlinear media. Furthermore, even the realization of plane waves may be practically impossible due to instabilities ubiquitous for nonlinear systems.

Here, we show that all the above challenges can be overcome: CPA for nonlinear waves does exist, can be observed experimentally,  and even can be more easily achieved due to the intrinsic nonlinearity. 
Theoretical indication for such CPA stems from the existence of stable constant-amplitude currents in a nonlinear waveguiding circle with equal absorbing and lasing potentials
~\cite{ZK}.  
Experimental indication comes from recent experiments  
on driven dissipative Josephson systems~\cite{Labouvie16}.   

 \section*{Results}

In this work, we consider \red{an atomic Bose-Einstein condensate (BEC) residing in a periodic potential, realized by an optical lattice. The superfluid nature of the Bose-Einstein condensate allows for tunneling between the wells, while interatomic collisions lead to an intrinsic nonlinearity.} One of the wells is rendered \red{absorptive} by applying an electron beam~\cite{Ott}, which removes atoms \red{from that well}. \red{The effective experimental system is sketched in Fig.~\ref{fig:working_principle}B: a well with linear absorption, embedded between two tunneling barriers, is coupled at both ends to a nonlinear waveguide.} \red{For an introduction to the experimental techniques for manipulating ultracold atoms in optical potentials, the reader is referred to Ref.\,\cite{Weidemuller}. Experimental details of the optical lattice, the preparation of the BEC and the experimental sequence are given in the Supplementary material.} The depth of the periodic potential and  number of atoms $N$ in each lattice site ($N\approx 700$) \red{are chosen such that we can apply} the tight-binding approximation of mean-field dynamics, when the condensate is described 
in terms of the density amplitudes $\psi_n(t)$~\cite{TrombSmerzi,AKKS}:   
\begin{eqnarray}
\label{DNLS}
i\hbar\frac{d\psi_n}{dt}=-J\left(\psi_{n-1}+\psi_{n+1}\right)+U|\psi_n|^2\psi_n-i\frac{\hbar\gamma}{2}\psi_n \delta_{n0}.
\end{eqnarray}
Here, $n$ enumerates the lattice sites and  $\gamma\geq 0$ describes the dissipation strength applied to the site $n=0$. Theoretically, such a system was first considered in~\cite{BKPO} within the framework of the Gross-Pitaevskii equation. Later on, it was treated within the Bose-Hubbard model~\cite{Barmettler2011,Witthaut2011}. Current states in BECs in the presence of dissipation and external drive were also studied theoretically in  \cite{ZKBO,ZK} and experimentally in  \cite{Barontini2013,Eckel2014,Labouvie16}.

Since CPA is a stationary process, we look for steady-state solutions of (\ref{DNLS}) in the form $\psi_n(t)=e^{-i(\mu/\hbar) t}u_n$, where all $u_n$ are time independent and $\mu$ is the chemical potential. First, we revisit the linear case corresponding to non-interacting atoms:
$\tilde{\mu} \tilde{u}_n=-J\left(\tilde{u}_{n-1}+\tilde{u}_{n+1}\right)-i(\hbar \gamma/2)\tilde{u}_n \delta_{n0}$, where in order to emphasize  the limit $U=0$ we use tildes. 
In the absence of dissipation, i.e., at $\gamma=0$, the dispersion relation in the tight-binding approximation reads $\tilde{\mu}=-2J\cos q$, where $q\in [0,\pi]$ is the wavenumber.
When dissipation is applied at $n=0$, we consider the left $\tilde{u}_n^L=a^Le^{iqn}+b^Le^{-iqn}$ for $n\leq -1$ and right $\tilde{u}_n^R=a^Re^{iqn}+b^Re^{-iqn}$ for $n\geq 1$ solutions, where $a^{L}$ and $b^{R}$ ($a^{R}$ and $b^{L}$) are the incident (reflected) waves from left ($L$) and right ($R$), respectively 
(see Fig.~\ref{fig:working_principle}). The transfer $2\times2$ matrix ${\bf M}$ with the elements $M_{ij}(q)$, is defined by the relation $(a^R,b^R)^T={\bf M}\,(a^L,b^L)^T$, where $T$ stands for transpose. 
Computing  ${\bf M}$ 
(see Methods),   one verifies that for  $q=q_\star$ and $q=q_\star^1=\pi-q_\star$ where
\begin{eqnarray}
\label{q_star}
q_\star=\arcsin\left(\frac{\hbar\gamma}{4J}\right),
\end{eqnarray}
the element $M_{11}$ vanishes: $M_{11}(q_\star)=0$, while the other elements become $M_{21}(q_\star)=-M_{12}(q_\star)=1$ and $M_{22}(q_\star)=2$. At these wavenumbers the problem admits a solution consisting of only incident waves, i.e.  $a^L=b^R$ and $b^L=a^R=0$. Thus, two CPA-states 
occur for slow ($q=q_\star$) and fast ($q=q_\star^1$) matter waves. 
The points $q_\star$ and $q_\star^1$ are called {\em time-reversed spectral singularities}.   

If instead of eliminating, one coherently injects atoms into the site $n=0$, i.e. $\gamma<0$, model (\ref{DNLS}) admits {\em spectral singularities} $\bar{q}_\star=-q_\star$ and  $\bar{q}_\star^1=-q_\star^1$ at which   $M_{22}(\bar{q}_\star)=0$ while the other elements become $M_{12}(\bar{q}_\star)=-M_{21}(\bar{q}_\star)=1$ and $M_{11}(\bar{q}_\star)=2$. Now, the solution  $a^R=b^L$ and $a^L=b^R=0$ describes coherent wave propagation outside the ``active'' site, corresponding to a matter-wave laser. Since the change $\gamma\to-\gamma$ in Eq.~(\ref{DNLS}) is achieved by applying the  Wigner time reversal operator $\mathcal{T}$: $\mathcal{T}\Psi({\bf r},t)=\Psi^*({\bf r},-t)$,  
a coherent perfect absorber corresponds to a time-reversed laser~\cite{StonePRL}.

The CPA solutions of model (\ref{DNLS}) for linear waves have the following properties: It exists only for dissipation rates with $\gamma\leq\gamma_{\rm th}=4 J/\hbar$. The amplitude of the  absorbed waves is constant in all sites, including the site where atoms are eliminated, and 
the group velocity at $q_\star$ is directly set by the decay rate: 
$v_{\rm g}(q_\star)=d\tilde{\mu}/dq|_{q_\star}=2J\sin q_\star=\hbar \gamma/2$. 

Bearing these properties in mind, we now turn to the nonlinear problem, setting $U>0$ (repulsive interactions among the atoms). We search for a steady-state  solution of (\ref{DNLS}) with a constant amplitude $\rho$ in each lattice site. The requirement for the existence of only left- and right- incident waves can be formulated as $u_n^L=\rho e^{iqn}$ for $n\leq -1$, $u_n^R=\rho e^{-iqn}$ for $n\geq 1$, and $u_0=\rho$. {This fixes   $\mu=-2J\cos q+U\rho^2$,}
and the  matching conditions  
at $n=0$ imply  
that the steady-state solution exists only if $q=q_\star$ (and $q=q_\star^1$)  given by  
(\ref{q_star}). Thus, we have obtained {\it CPA for nonlinear matter waves}, which still corresponds to the time-reversed laser. In addition, as in the linear case, replacing the dissipation with  gain  (i.e., inverting the sign of  $\gamma$), one obtains the constant amplitude outgoing-wave solution:  $u_n^L=\rho e^{-iq_\star n}$ for $n \leq -1$, $u_n^R=\rho e^{iq_\star n}$ for $n\geq 1$.

One essential difference between linear and nonlinear CPA  
is particularly relevant for the experimental observation of the phenomenon: the stability of the incoming superfluid currents. The stability analysis (see Methods) 
shows
that the nonlinearity qualitatively changes the result: only slow currents ($q=q_\star$) can be perfectly absorbed, while the fast nonlinear currents ($q=q_\star^1$) are  
dynamically unstable.

\begin{figure}
	\includegraphics[width=\columnwidth]{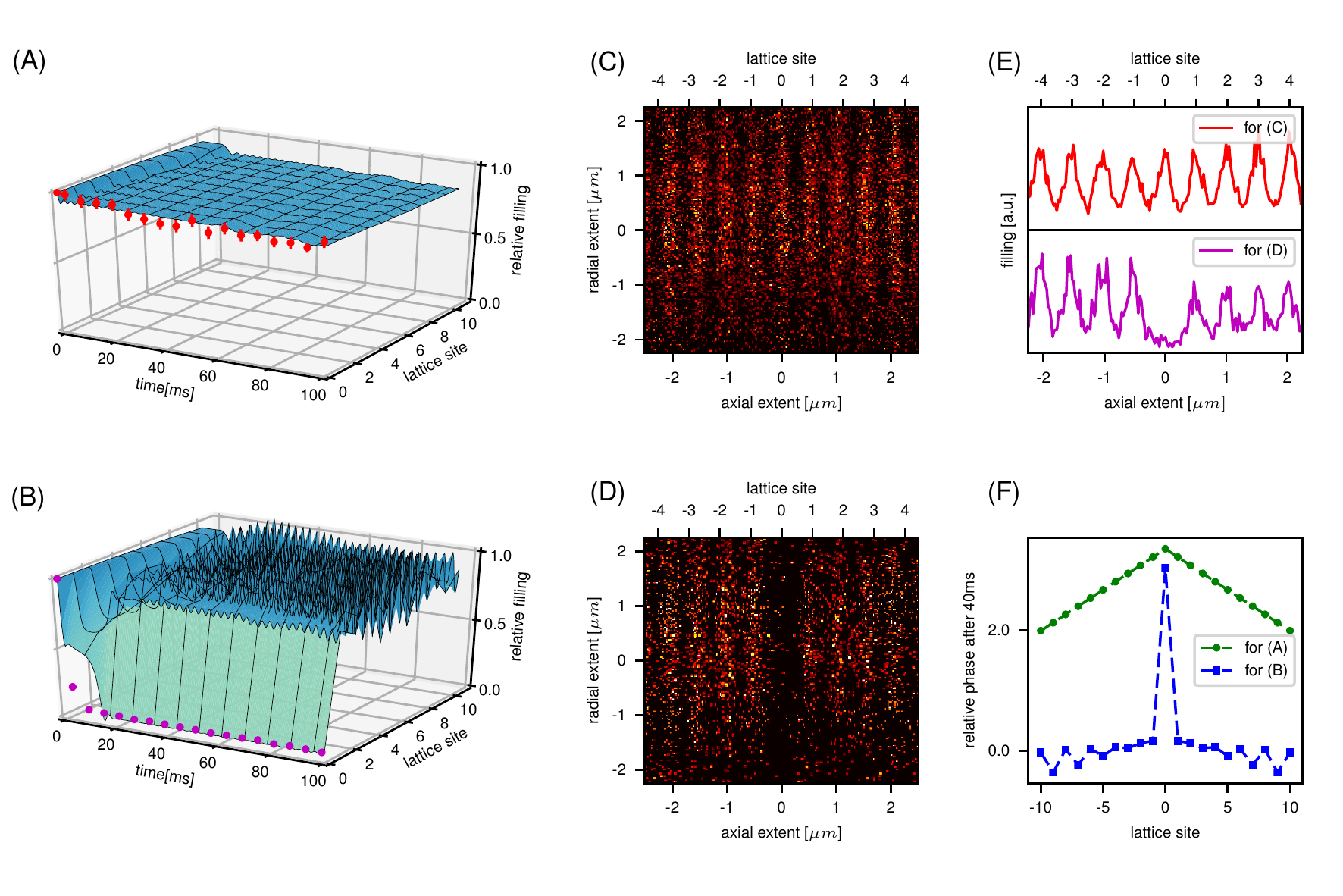}
	\caption{Coherent perfect absorption in a Bose-Einstein condensate residing in an optical lattice. At $t=0$, the elimination of atoms from the lattice site $n=0$ starts. (A), (A) Time evolution of the lattice filling in the CPA regime ($\gamma\approx125\,$s$^{-1}$, (A)) and above the breaking point ($\gamma\approx1000\,$s$^{-1}$, (A)). Experimental data are shown as red points, the numerical simulations are shown in blue. For clarity, we only show the first ten sites on one side of the system. In (A), the density remains uniform across the lattice, which is the signature of CPA. In (B), the density at the dissipative site drops off rapidly, and no CPA can be observed. (C),(D) Corresponding images of the atomic distribution in the lattice after the sequence. (E) Integrated atomic density of (C) and (D). (F) Unwrapped radian argument $\arg\psi_n$  at $t=40$ms for the solution shown in (A) [green curve] and (B) [blue curve]. For all panels, we have   $U/\hbar=2600$\,$s^{-1}$, $J/\hbar=229$\,$s^{-1}$. The error bars in (A) and (B) indicate the statistical error, resulting from the summation over 50 experimental runs.}
	\label{fig:num}
\end{figure}

The CPA solution is mathematically valid only in the infinite lattice, because the absorption at the center must be compensated by steady particle fluxes incoming from infinity. However, the CPA phenomenon is structurally robust and can be observed as a   quasi-stationary regime even in a  finite lattice. To demonstrate this, we simulated numerically Eq.~(\ref{DNLS}) with about 200 sites. The  initial condition corresponds to the ground state of a BEC with an additional small harmonic confinement along the lattice direction.  
Figures~\ref{fig:num}A and~\ref{fig:num}B show the obtained behavior for  dissipation strengths $\gamma$ below (a) and above (b) the CPA-breaking point $\gamma_{\rm th}$.  
Fig.~\ref{fig:num}A shows that  
the  solution rapidly enters a quasi-stationary regime where its density remains constant in space and is only weakly decaying in time due to an overall loss of atoms in the system. Above the breaking value 
(Fig.~\ref{fig:num}B),
a strong decay sets in and the atomic density is not homogeneous in space any more. An important feature of CPA is the balanced superfluid currents towards the dissipative site, characterized by the distribution of phases, illustrated in 
Fig.~\ref{fig:num}C. The CPA regime manifests itself in the $\Lambda$-shaped phase profile whose slope is $q_\star$ for negative $n$  and $-q_\star$ for positive $n$. This phase pattern is completely different when the system is not in the CPA regime: it is nearly constant, showing weak nonmonotonic behavior for positive and negative $n$, with a large jump at the central site.

Together with the numerical simulation, we also show in 
Fig.~\ref{fig:num}A the corresponding experimental results. The experimentally measured filling level of the dissipative site shows very good agreement with the numerical simulations: the atom number in the dissipated site is constant in time and equal to all neighboring sites. This steady state is the experimental manifestation of CPA of matter waves. The CPA solution is established also for other values of the dissipation strength. This highlights the fact that the nonlinearity together with the dissipation generates an effective attractor dynamics towards the CPA solution. Increasing the dissipation above a critical value leads to a qualitative change in the behaviour 
(Fig.~\ref{fig:num}B). 
In accordance with the theoretical prediction, the occupation in the dissipated site rapidly drops and stays small. Hence, CPA can indeed only be observed in a finite parameter window.

The theoretical results predict the transition from CPA and the non-CPA regime at $\gamma_c=4J/\hbar$, above which no quasi steady-state is established anymore.  In the experiment, a qualitatively similar situation occurs 
(Fig.~3). 
However the CPA regime breaks down  at a lower dissipation rate of $\gamma_{\mathrm exp}\approx J/\hbar$.
This can be explained by two factors. First, the transverse extension of each lattice site, not fully accounted for by the tight-binding approximation (\ref{DNLS}), makes the condensate vulnerable against transverse instabilities~\cite{Herring} which can develop at smaller $\gamma$ than predicted by (\ref{DNLS}). The second factor is related to the way the experiment is conducted. At $t=0$ the condensate is loaded in a lattice and is characterized by the chemical potential, which is determined by the trap geometry and by the filling of each lattice site.  When the elimination of atoms starts, the system is induced in an unstable regime. Quasi-stationary behaviour is only possible if the chemical potential is not changed appreciably under the action of the dissipation. This requires the filling of the central site due to tunneling from the neigbouring ones to be  fast enough to compensate the loss of atoms. Thus the tunneling time estimated as $\tau_{\rm tun}\sim \hbar/J$ should be of the order of or smaller than the inverse loss rate $\gamma^{-1}$, otherwise the induced thermodynamical equilibrium at $t=0$ cannot be compensated by the incoming superfluid currents and the collective dynamics described by Eq.~(\ref{DNLS}) cannot be established. This gives an estimate $\gamma\sim J/\hbar$ for the threshold dissipation rate.

\begin{figure}
	\centering
	
\includegraphics[width=0.6\columnwidth]{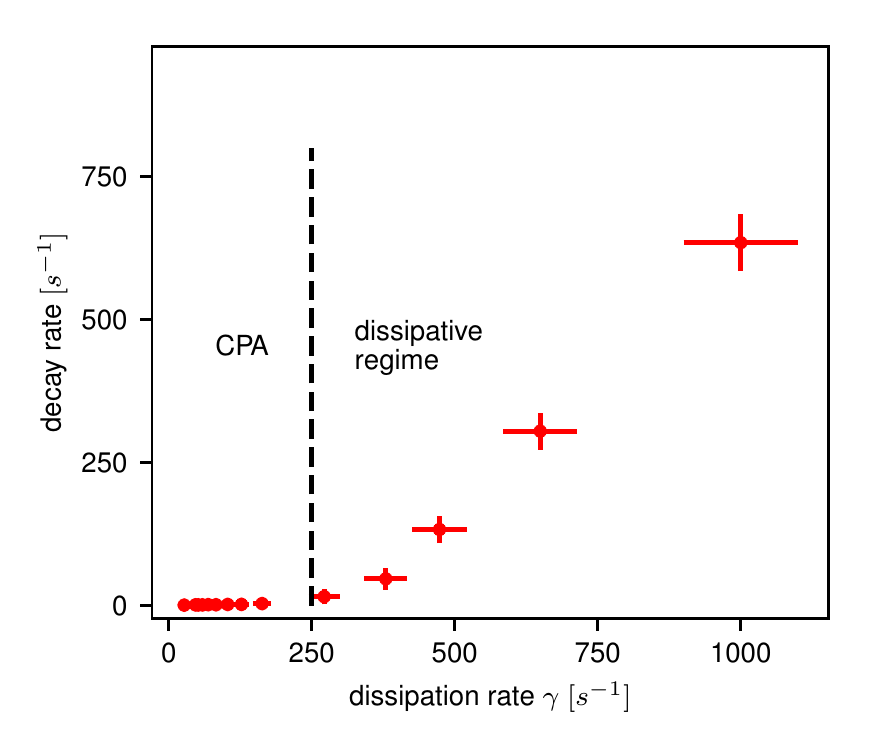}%
\caption{Experimentally measured decay rate of the filling level. Up to a critical dissipation
	strength $\gamma_{exp}\approx$250\,s$^{-1}$ (black dashed line), the filling level remains constant (compare Fig.~2A), corresponding to the CPA regime. Above this value, the dissipation dominates the dynamics and the filling level decays exponentially (compare Fig.~2B). The statistical error of the decay rate is smaller than the size of the points, however, we estimate a $5\%$ systematic error due to technical imperfections like drifts of the electron beam current. The error in the dissipation rate originates from the calibration measurement (see Supplementary Materials).}
\label{fig:CPA_transition}
\end{figure}

 \section*{Conclusion}
Our results present the
Proof-of-Concept of the CPA paradigm \red{nonlinear waves. The  experimental setting explored here} can be straightforwardly generalized to BECs of other types like spin-orbit coupled, fermionic, and quasi-particle ones, and furthermore to other branches of physics, including nonlinear optics of Kerr media and acoustics.  Our system can also be exploited as a platform for studying \red{superfluid flows in a linear geometry (which is aletrnative to most commonly used annular traps), as well as for understanding} the fundamental role of Bogoliubov phonons in stabilizing quantum states.  \red{Since CPA can be viewed as time-reversed  lasing, the reported experimental results pave the way to implementing a laser for matter waves, for which elimination of atoms from the central site should be replaced by loading atoms. Furthermore, the observation of CPA in nonlinear media, and possible lasing of matter waves, can be viewed as an additional element for rapidly developing area of quantum technologies based on atomtronics~\cite{atomtronics}. The reported results also open a possibility of using CPA regimes in nonlinear optical circuits.} 

In the general context of scattering by dissipative potential~\cite{Muga}, given the fact, that the atomic interactions can be tuned by a Feshbach resonance, a linear spectral singularity can be experimentally realized by starting from the nonlinear case and subsequently reducing the interactions  to zero adiabatically. Such a scenario explicitly exploits the attractor nature of the CPA solution.  
\red{
	  Being an attractor in essentially nonlinear system, CPA can serve as a mechanism to  control  parameters of superfluid flow parameters, such as the chemical potential, superfluid velocity, or sound velocity, in a particularly simple way.}

 \section*{Materials and Methods}

\subsection*{The transfer matrix}

To compute the transfer matrix ${\bf M}$, we denote the solution in the point $n=0$ by $u_0$ and consider 
the equation 
\begin{eqnarray}
\tilde{\mu} \tilde{u}_n=-J\left(\tilde{u}_{n-1}+\tilde{u}_{n+1}\right)-i\frac{\hbar \gamma}{2}\tilde{u}_n \delta_{n0}, 
\end{eqnarray}
describing stationary currents, in the points $n=0$ and $n=\pm 1$ using the explicit forms  for the waves in the left $\tilde{u}_n^L=a^Le^{iqn}+b^Le^{-iqn}$ ($n\leq -1$) and the right $\tilde{u}_n^R=a^Re^{iqn}+b^Re^{-iqn}$ ($n\geq 1$) half-space.  
From the equation with $n=0$ and using the expression for the chemical potential $\tilde{\mu}=-2J\cos q$, we obtain $u_0$:
\begin{eqnarray}
u_0=\frac{2J}{4J\cos q-i\hbar\gamma}\left(a^Le^{-iq}+b^Le^{iq}+a^Re^{iq}+b^Re^{-iq}\right).
\end{eqnarray}
With this expression, the equations at $n=\pm1$ are transformed to a linear algebraic system, which is solved for the pair $(a^R,b^R)$, giving their expressions through $(a^L,b^L)$, thus determining the transfer matrix:
\begin{equation}
M_{11} =   \frac{4J\sin q - \hbar\gamma}{4J\sin q}, \quad M_{12} = - \frac{ \hbar\gamma}{4J\sin q}, \quad
M_{21} = \frac{ \hbar\gamma}{4J\sin q}, \quad M_{22} =  \frac{4J\sin q + \hbar\gamma}{4J\sin q}.
\end{equation}

\subsection*{Stability analysis}

We analyze the stability for a BEC within the framework of the discrete model 
\begin{eqnarray}
\label{DNLS_SM}
i\hbar\frac{d\psi_n}{dt}=-J\left(\psi_{n-1}+\psi_{n+1}\right)+U|\psi_n|^2\psi_n-i\frac{\hbar\gamma}{2}\psi_n \delta_{n0},
\end{eqnarray}
and require that the left and right incident superfluid currents have to be stable. Their stability is determined by the stability of the corresponding Bogoliubov phonons on an infinite homogeneous lattice (i.e., without applied removal of atoms). The stability of the homogeneous lattice   is found using the substitution 
\begin{eqnarray}
\label{Bogoliubov1}
\psi_n(t)=e^{-i(\mu/\hbar) t+i qn}\left(\rho+v_n e^{-i\omega t+ikn}+w_n^{*}e^{i\omega^* t-ikn}\right),
\end{eqnarray}
where $\rho>0$ characterizes the   uniform density, and $|v_n|\,, |w_n|\ll \rho$ are small perturbations. Linearizing  Eq.~(\ref{DNLS_SM}) (with $\gamma=0$) with respect to $v_n$ and $w_n$, we find two dispersion  branches: 
\begin{eqnarray}
\hbar \omega_\pm (q,k)=2J\sin(k)\sin(q)\pm 2\sin\left(\frac{k}{2}\right)
\sqrt{2J\cos(q)  \left[U\rho^2+2J\cos(q)\sin^2\left(\frac{k}{2}\right)\right]}.
\end{eqnarray}
Consider now a positive scattering length, $U>0$, which corresponds to the experiments reported here. One can then identify the stability domain for Bogoliubov phonons, and hence the stability of the superfluid current, requiring   $\omega_\pm$ to be  real for the given $q$ and all real $k$. This results in the constraint  $0\leq q<\pi/2$, i.e. only slow currents are dynamically stable. 

\subsection*{Experimental setup}

We use a Bose-Einstein condensate (BEC) of $^{87}$Rb with about $45\times10^3$ atoms in a single beam dipole trap realized by a CO$_2$-laser \red{(maximum power $10\,$W, beam waist $30\,\mu$m}). \red{The condensate is cigar-shaped and has dimensions of $80\,\mu$m $\times$ $6\,\mu$m $\times$ $6\,\mu$m.} We then load the BEC into a one-dimensional optical lattice created by two blue detuned laser beams ($\lambda=774\,$nm, \red{beam waist $500\,\mu$m}) crossed at an angle of $90{^\circ}$. \red{The linear polarization of both laser beams is along the same direction, such that the interference pattern is maximally modulated}. The resulting lattice has a period of $d=547\,$nm. The trap-frequencies in a lattice site are $\nu_r=165\,$Hz (transverse direction) and $\nu_{z}=12\,$kHz (lattice direction). Each site contains a small, pancake-shaped BEC with about 700 atoms (value in the center of the trap). The total number of lattice sites is about 200. The lattice depth $V_0$ in units of the recoil energy $E_r=\pi^2\hbar^2/(2md^2)$ ($m$ is the mass of the atom) is given by $V_0=10E_r$. \red{An electron column, which is implemented in our experimental chamber, provides a} focused electron beam, which is used to introduce a well-defined local particle loss as a dissipative process in one site of the lattice. In order to ensure a homogeneous loss process over the whole extension of the lattice site, we rapidly scan the electron beam in the transverse direction (3\,kHz scan frequency) with a sawtooth pattern. To adjust the dissipation strength $\gamma$, we vary the amplitude of the scan pattern. \red{An image of the experimental chamber together with a sketch of the optical trapping configuration is provided in the Supplementary Material (S1).}

\subsection*{Why the lattice is necessary}

As it is mentioned in the main text the superfluid currents under localized dissipation were studied previously in inhomogeneous BECs~\textit{(22,25)}, where no CPA was observed. Mathematical solutions of the Gross-Pitaevskii equation with localized dissipation describing such models can however be found. Such solutions are stable and have stationary amplitudes. Indeed, consider the stationary Gross-Pitaevskii equation with strongly localized dissipation modeled by the Dirac delta-function $\gamma_0\delta(x)$, where $\gamma_0$ is a positive constant, without any optical lattice  
\begin{eqnarray}
\label{eq:cont}
\mu\psi=-\psi_{xx}-i\gamma_0\delta(x)\psi+g|\psi|^2\psi.
\end{eqnarray}  
Here $g>0$. One can verify, that the function 
\begin{equation}
\label{eq:contCPA}
\psi =\rho_0 e^{-i\gamma_0 x   {\rm sign}(x) /2}, 
\end{equation} 
is a solution of Eq.~(\ref{eq:cont}) with the chemical potential $\mu = g\rho_0^2 + \gamma_0^2/4$.

This raises questions about the role of the optical lattice and about its  necessity for realizing CPA experimentally. The answer resides in the way of exciting the CPA regime. A strictly homogeneous background density can only exist if the dissipation is point-like (described by the Dirac delta-function  $\delta$) and thus experimentally unrealistic. Any finite size, even very narrow, dissipation generates Bogoliubov phonons at the instant it is applied. In the continuous model, the phonons can propagate with arbitrary group velocity, contrary to the lattice described in the tight-binding model. Switching on dissipation therefore induces an extended domain of the condensate in a dynamical regime, and fast matter waves propagating outwards the dissipation domain cannot be stabilized by the incoming flows. Thus, the lattice, on the one hand creates conditions where the dissipation is effectively point-like (i.e. applied to a single cell) and on the other hand limits the group velocity of the phonons, allowing for establishing the equilibrium state.

\subsection*{Details of numerical simulations}

In the numerical simulations, we used the following model which corresponds  to the Gross-Pitaevskii equation from the main text with an additional weak parabolic  confinement $\alpha n^2\psi_n$ which models the optical dipole trap potential used in the experimental setup:
\begin{equation}
\label{eq:GPEn2}
i\hbar \frac{d\psi_n}{dt}= - J (\psi_{n-1} + \psi_{n+1}) + U|\psi_n|^2 \psi_n - \frac{i\gamma\hbar}{2}\delta_{n0}\psi_n + \alpha n^2\psi_n.
\end{equation}
The coefficient $\alpha$  determines the strength of the parabolic trapping and  amounts to
\begin{equation}
\alpha = \frac{m\omega^2 d^2}{2\hbar }   = 0.98~\mbox{s}^{-1},
\end{equation}
where $m=1.44 \cdot 10^{-25}$~g is the mass of the atom, $d=547$~nm is the lattice period, and $\omega=2\pi \cdot 11$~Hz is the axial trapping frequency of the dipole trap. 
As in the main text, for other parameters we have 
$J/\hbar = 229~\mbox{s}^{-1}$ and $ U/\hbar =2600~\mbox{s}^{-1}$.

To convert the equation (\ref{eq:GPEn2}) in the form suitable for numerical calculations, we divide each term by $J$ and introduce the ``new time''  $\tau = (J/\hbar) t$ which transforms (\ref{eq:GPEn2}) to
\begin{equation}
\label{eq:numerical}
i\frac{d\psi_n}{d\tau}= - (\psi_{n-1} + \psi_{n+1}) + \tilde{U} |\psi_n|^2 \psi_n - \frac{i\tilde{\gamma}}{2}\delta_{n0}\psi_n + \tilde{\alpha} n^2\psi_n,
\end{equation}
where
$\tilde{U} =U/J\approx  11,$ $\tilde{\alpha} =\alpha/ J\approx 0.004$, and $\tilde{\gamma} = \gamma \hbar / J$.

For $\tilde{\gamma}=0$, Eq.~(\ref{eq:numerical}) has  an approximate stationary  ground-state Thomas-Fermi  solution 
\begin{eqnarray}
\label{eq:TF}
\psi_n = e^{-i( \tilde{U}\rho^2-2) t} w_n, \qquad 
w_n=\left\{\begin{array}{ll}
\sqrt{\rho^2 - \tilde{\alpha} n^2/\tilde{U}}, &  |n|<N_{TF}\\
0,& |n|>N_{TF}.
\end{array}
\right.
\end{eqnarray}
where the ``discrete Thomas-Fermi radius''  $N_{TF}$ is determined by the condition $\tilde{U}\rho^2 - \tilde{\alpha} N_{TF}^2= 0$, and $\rho^2=1$ is the normalized background density, i.e., $|w_n|^2\approx 1$ in the central region. In our   case $N_{TF} \approx 50$.

We solve equation (\ref{eq:numerical})  for $\psi_n(\tau)$ on the grid  of 201 sites $n=-100, \ldots, 100$, where $n=0$ corresponds to the site with the losses, and subject to the zero boundary conditions $\psi_{-100}(\tau) =\psi_{100}(\tau)=0$. 

For the initial condition,  we used the ground state Thomas-Fermi distribution from (\ref{eq:TF}): $\psi_n(t=0)=w_n$. As discussed in the main text, in the parametric range corresponding to the existence of the CPA, the initial condition rapidly evolves to the quasi-stationary CPA solution characterized by the almost uniform density in the central region ($n=-10,\ldots,10$), whereas in the absence of the CPA-regime the initial condition rapidly develops    strong instability in the central region.


\bibliography{scibib}

\bibliographystyle{Science}

\section*{Acknowledgments}
 \subsection*{Funding:}
 V.V.K. was supported by the FCT (Portugal) under Grant No. UID/FIS/00618/2013. H.O. and A.M. acknowledge financial support by the DFG within the SFB/TR 185. C.B., J.B. and R.L. acknowledge financial support by the DFG within the SFB/TR 49. C.B. and R.L. acknowledge financial support by the MAINZ graduate school. D.A.Z. was financially supported by Goverment of Russian Federation (Grant 08-08).

\subsection*{Author contributions:}
H.O. and V.V.K. conceived the idea.  R.L. and B.S. performed the experiment. R.L. and A.M. analysed the data. H.O. supervised the experiment.  V.V.K. developed the analytical model.  D.A.Z. performed the numerical simulations. V.V.K., H.O. A.M. and D.A.Z. prepared the manuscript. All authors contributed to the data interpretation and final manuscript preparation.

\subsection*{Competing interests:} Authors declare no competing interests.

\clearpage

\newpage	
	
	\begin{figure}		
		\centering
		
\includegraphics[width=0.8\columnwidth]{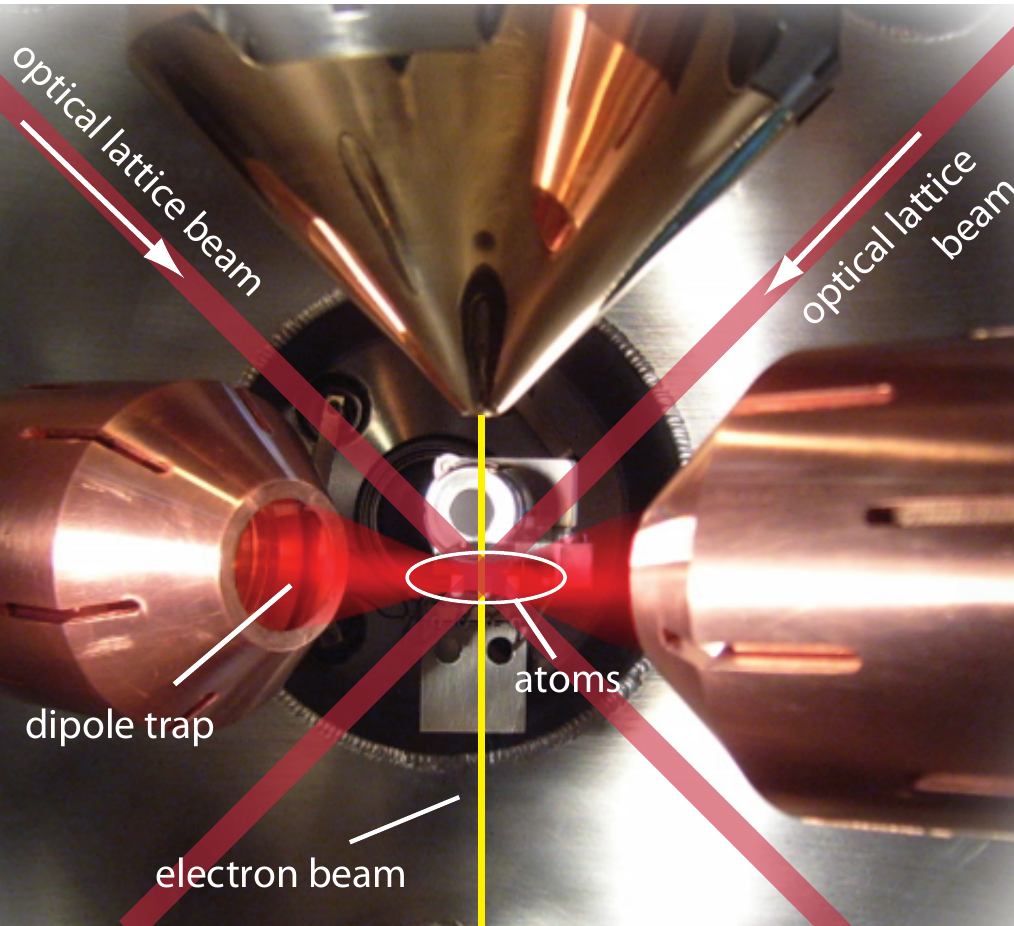}%
\end{figure}

\noindent {\bf Fig. S1.} Photograph of the vacuum chamber and sketch of the optical trapping scheme. The Bose-Einstein condensate is prepared in a single beam optical dipole trap. The periodic potential is created by two interfering optical lattice beams, which create a standing light wave at the position of the atoms. The combination of both potentials results in the physical situation as depicted in Fig.\,1B. An electron column provides a focused electron beam, which is pointed at one of the potential wells. Scattering processes between the electrons and the ultracold atoms lead to local loss (absorption) of the atoms. The size of the image section is about 15\,cm $\times$ 15\,cm.

\end{document}